\documentclass[10pt]{article}
\usepackage[OE]{express}
\usepackage{amsmath}
\usepackage{braket}
\usepackage{color,soul}
\usepackage{float}
\usepackage{multirow}
\usepackage{amssymb}
\usepackage{array}

\begin{document}

\title{50-GHz-spaced comb of high-dimensional frequency-bin entangled photons from an on-chip silicon nitride microresonator}

\author{Poolad Imany,\authormark{1,2,6,*} Jose A. Jaramillo-Villegas,\authormark{1,2,3,6} Ogaga D. Odele,\authormark{1,2} Kyunghun Han,\authormark{1,4} Daniel E. Leaird,\authormark{1,2} Joseph M. Lukens,\authormark{5} Pavel Lougovski,\authormark{5} Minghao Qi,\authormark{1,4} and Andrew M. Weiner\authormark{1,2,4}}

\address{\authormark{1}School of Electrical and Computer Engineering, Purdue University, West Lafayette, IN, USA\\
\authormark{2}Purdue Quantum Center, Purdue University, West Lafayette, IN, USA\\
\authormark{3}Facultad de Ingenier\'{i}as, Universidad Tecnol\'{o}gica de Pereira, Pereira, RIS, Colombia\\
\authormark{4}Birck Nanotechnology Center, Purdue University, West Lafayette, IN, USA\\
\authormark{5}Quantum Information Science Group, Oak Ridge National Laboratory, Oak Ridge, TN, USA\\
\authormark{6}These authors contributed equally to this work\\
}

\email{\authormark{*}pimany@purdue.edu} 


\begin{abstract}
Quantum frequency combs from chip-scale integrated sources are promising candidates for scalable and robust quantum information processing (QIP). However, to use these quantum combs for frequency domain QIP, demonstration of entanglement in the frequency basis, showing that the entangled photons are in a coherent superposition of multiple frequency bins, is required. We present a verification of qubit and qutrit frequency-bin entanglement using an on-chip quantum frequency comb with 40 mode pairs, through a two-photon interference measurement that is based on electro-optic phase modulation. Our demonstrations provide an important contribution in establishing integrated optical microresonators as a source for high-dimensional frequency-bin encoded quantum computing, as well as dense quantum key distribution.
\end{abstract}


\bibliographystyle{osajnl}
\bibliography{FBEMRR}

\section{Introduction}
Quantum information processing (QIP) has gained massive attention in recent years as it promises to solve some exponentially hard problems in polynomial time through quantum computation \cite{nielsen2010quantum, steane1998quantum, walther2005experimental}, as well as having other unique capabilities such as fully secure communications through quantum key distribution \cite{shor2000simple,barreiro2008beating,gisin2007quantum,ali2007large,zhong2015photon}, and enhanced sensing through quantum metrology \cite{giovannetti2006quantum}. Typical QIP systems are based on two-level quantum states, also called qubits. To simplify the complexity of quantum circuits \cite{knill2001scheme,o2003demonstration} and increase the practicality of quantum computation, high-dimensional entangled states (entangled qudits) are strong candidates as a result of their robustness and stronger immunity to noise, compared to two-dimensional systems \cite{babazadeh2017high,lanyon2009simplifying,xie2015harnessing,sheridan2010security}.

In photonics, amongst different degrees of freedom capable of high-dimensionality, the frequency domain---using single or entangled photons in a coherent superposition of multiple frequency bins---offers both compatibility with fiber transmission and more robust and scalable systems because it does not require stabilization of interferometers or complex beam shaping \cite{xie2015harnessing,ramelow2009discrete}. But while frequency-bin entangled photons (also referred to as biphoton frequency combs or BFCs) have been explored through spontaneous parametric down-conversion (SPDC) together with cavity and programmable spectral filtering \cite{lu2003mode,olislager2010frequency,bernhard2013shaping}, the bulk platform is faced with the drawback of low scalability and high cost. To overcome these disadvantages, integrated optical microresonators offer a solution that is highly compatible with semiconductor foundries. Such chip-scale devices have been used to generate entangled photons with a comb-like spectrum \cite{reimer2016generation,mazeas2016high,jaramillo2017persistent}. Time-bin entanglement for a single comb line pair from microresonators has been verified in \cite{reimer2016generation,mazeas2016high,ramelow2015silicon}, and in \cite{jaramillo2017persistent} time-bin entanglement was demonstrated for multiple comb line pairs simultaneously. Yet these studies did not show the ensuing photon states to be in a coherent superposition of multiple frequency-bins. The difficulty of this measurement stems from the large Free Spectral Range (FSR) of conventional microring resonators (typically a few hundred GHz) which results in temporal correlation trains with periods of order several picoseconds, much faster than the timing resolution of standard single-photon detectors ($\sim$100 ps). As a result, direct detection of the comb-like photon pairs is incapable of showing spectral phase sensitivity, a condition required to prove frequency-bin entanglement.

Phase modulation has been used to mix frequency states of biphotons generated by SPDC \cite{olislager2010frequency}; frequency-bin entanglement was tested through analysis of two-photon interference as a function of the phase modulation amplitude.  Here we use phase modulation to overlap sidebands from either two or three different comb line pairs which are preselected and adjusted for equal amplitude by a programmable pulse shaper. This provides an indistinguishable superposition of frequency states for two-photon interference measurements which prove phase coherence and frequency-bin entanglement in two dimensions (qubits) and three dimensions (qutrits) for a silicon nitride on-chip BFC. In contrast to \cite{olislager2010frequency}, our approach provides a close analog with Franson interferometry methods that have been widely used for characterization of time-bin entanglement \cite{reimer2016generation,mazeas2016high,jaramillo2017persistent,ramelow2015silicon}. We presented a subset of these results in \cite{imany2017demonstration}; earlier we demonstrated the feasibility of this approach in measurements involving entangled photons generated via SPDC in \cite{imany2017two}. A similar technique was developed independently and presented in \cite{kues2017chip}, exploring frequency-bin entanglement for a Hydex microring resonator with 200 GHz FSR. Our experiments explore a larger microresonator with $\sim$50 GHz FSR. Due to the denser resonance spacing, we are able to identify up to 40 frequency modes from the Joint Spectral Intensity (JSI), a factor of 4 higher than in \cite{kues2017chip}, which suggests substantial potential to push towards higher dimensionality. From a practical perspective, the more closely spaced resonances should provide a better match to the capabilities of practical phase modulator technology, allowing a greater number of frequency modes to be superimposed for future studies of higher dimensionality entanglement. Demonstration of frequency-bin entanglement is a major step in qualifying integrated biphoton frequency comb sources for applications in scalable high capacity quantum computation \cite{lukens2017frequency} and dense quantum communications \cite{mower2013high}.

\section{Experiments}

\begin{figure}[t]
\centering\includegraphics[width=\textwidth]{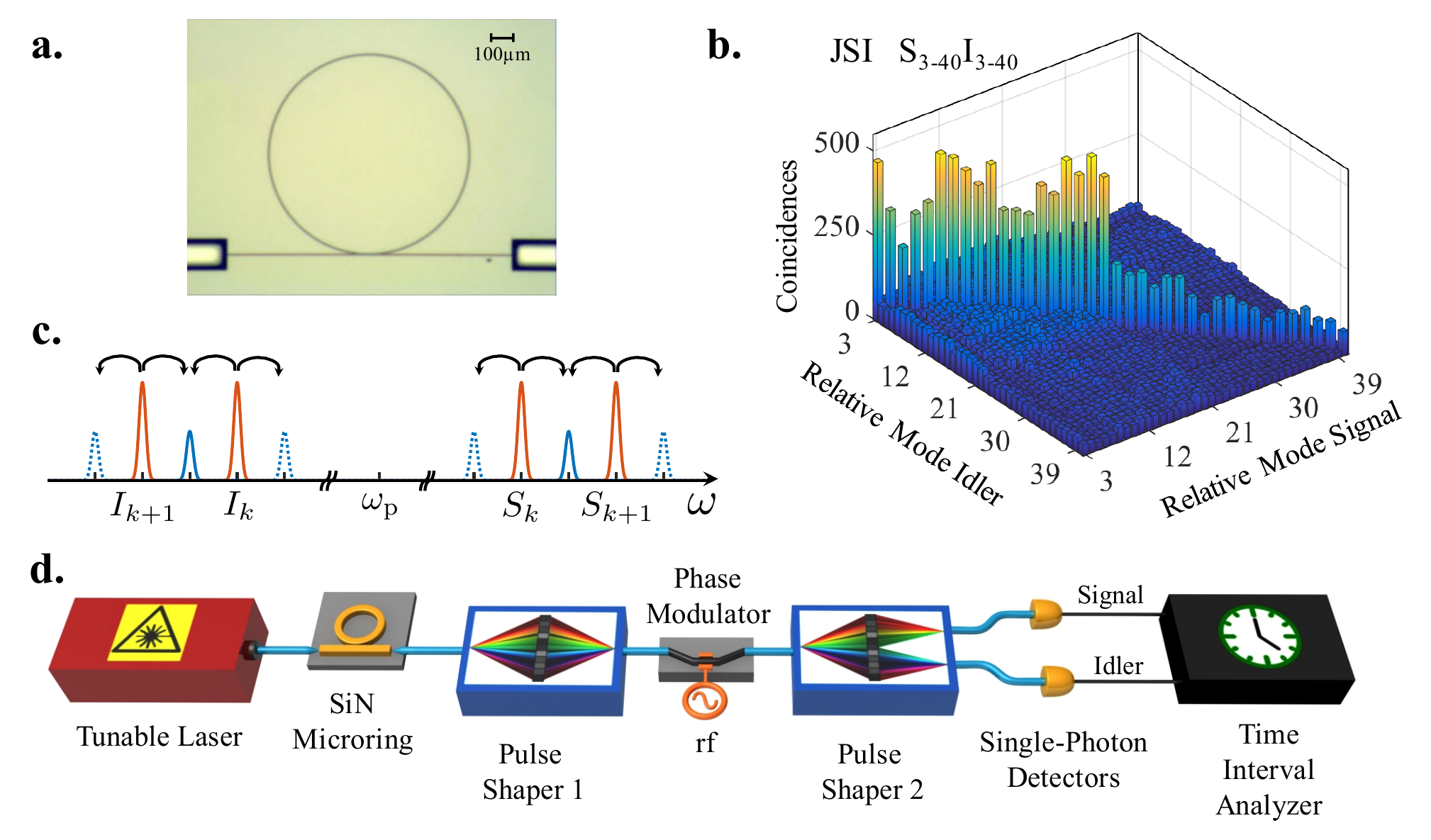}
\caption{(a) Microscope picture of the microring and U-grooves to support fiber coupling. (b) Joint spectral intensity for comb line pairs from 3 to 40. The background accidentals are not subtracted in this measurement and the coincidence to accidental ratio is about 10:1. (c) Illustration of biphoton spectrum after phase modulation. (d) Experimental setup.}
\label{Fig1}
\end{figure}

For our experiments, we use a silicon nitride microring resonator [Fig. \ref{Fig1}(a)] with a loaded quality factor of $\sim 2\times 10^6$ to generate entangled photons. The field possible within the microring corresponds to resonant modes with linewidths of $\sim$100 MHz and frequency separations just under 50 GHz. Hence when we pump the ring with a tunable continuous-wave laser (operating in the C-band), the spontaneous four wave mixing (SFWM) process leads to the generation of a quantum frequency comb with a frequency spacing and linewidth that mirror the resonance structure of the ring. Further details on the microring and experimental procedures are provided in Appendix A. Generally, the BFC state can be written as:

\begin{equation}
\label{Eq1}
\Ket{\Psi} = \sum_{k=1}^{N} { \alpha_{k} \ket{k,k}_{SI}}
\end{equation}

\begin{equation}
\ket{k,k}_{SI} = \int{d\Omega  \hspace{4pt} \Phi\left(\Omega - k\Delta\omega\right) \hspace{2pt} \Ket{\omega_{\textrm{\fontsize{6}{0}\selectfont{P}}}+\Omega,\omega_{\textrm{\fontsize{6}{0}\selectfont{P}}}-\Omega }_{SI}}
\label{Eq2}
\end{equation}

\noindent where $\ket{k,k}_{SI}$ represents the signal and idler photons from the $k^{\textrm{th}}$ comb line pair, $\alpha_k$ is a complex number describing the amplitude and phase of the $k^{\textrm{th}}$ comb line pair and $N$ is the total number of mode pairs, $\Phi(\Omega)$ is the lineshape function, $\Delta\omega$ is the FSR and $\omega_{\textrm{\fontsize{6}{0}\selectfont{P}}}$ is the pump frequency. The coherent superposition of $\ket{k,k}_{SI}$ states implied by Eq. (\ref{Eq1}) requires phase coherence between the frequency mode pairs, i.e., the different $\ket{k,k}_{SI}$ must be able to interfere.

\subsection{Joint spectral intensity}

We characterized the spectro-temporal correlations between combinations of frequency modes spanning a 38$\times$38 space (signal and idler lines 3--40). Using a programmable pulse shaper \cite{weiner2000femtosecond} as a tunable frequency filter, we route different signal and idler photons to a pair of single-photon detectors (SPDs) and record the relative arrival time of each photon pair with a Time Interval Analyzer (TIA). As expected, we observe tight temporal correlations only between energy matched comb lines spanning up to the 40$^{\textrm{th}}$ mode, as presented in the form of the JSI in Fig. \ref{Fig1}(b); the high diagonal coincidences reflect the energy matching in the SFWM process. The calculated lower bound of the Schmidt number for this JSI is $k_{\textrm{min}}=20$, which is a figure of merit for the degree of frequency correlations \cite{eckstein2014high}. Here we note that the JSI, unlike the joint spectral amplitude, lacks any phase information and cannot show phase coherence between different frequency mode pairs.

\subsection{Two-dimensional frequency-bin entanglement}

To show phase coherence between different comb line pairs, we implement the setup depicted in Fig. \ref{Fig1}(d). The output of the microring is coupled into pulse shaper 1, where in the first experiment we select only comb line pairs 6 (S$_6$I$_6$) and 7 (S$_7$I$_7$). Subsequently, we will use this pulse shaper to apply optical spectral phase to the comb lines. We also note that we use the first pulse shaper to equalize the contribution of the modes to coincidence counts. By doing so, we are making sure that $\left|\alpha_k\right|=\left|\alpha_{k+1}\right|$ for the rest of the experiments, which optimizes contrast in quantum interference. The selected lines are then coupled into an electro-optic phase modulator, which creates optical sidebands at frequency offsets equal to multiples of the radio frequency (rf) of the driving sinusoidal waveform, which we set to yield sidebands at half the spacing of the BFC [Fig. \ref{Fig1}(c)]. Then with pulse shaper 2, we pick out the sidebands which overlap midway between S$_6$-S$_7$ and I$_6$-I$_7$ [solid blue curves in Fig. \ref{Fig1}(c)], and route them to the SPDs and the TIA.

\begin{figure}[t]
\centering\includegraphics[width=7cm]{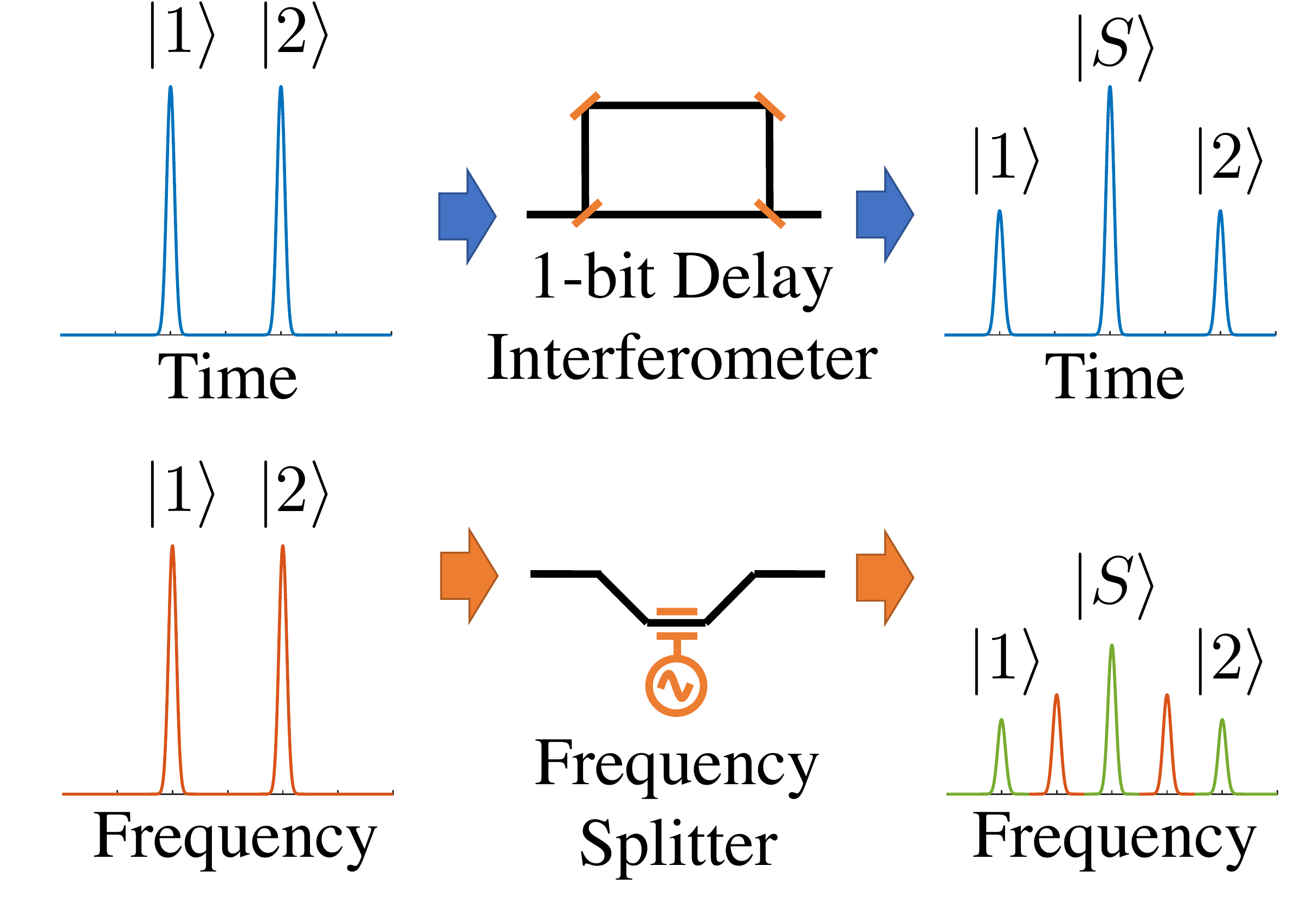}
\caption{Analogy between a 1-bit delay interferometer for forming projections of time-bin qubits and a frequency splitter for forming projections of a frequency-bin qubit. The green frequency bins after the frequency splitter are phase modulation sidebands from $\ket{1}$ and $\ket{2}$.}
\label{Fig2}
\end{figure}

Our frequency-bin entanglement verification scheme is a frequency domain analog of the Franson interferometry approach \cite{franson1989bell} widely used to verify time-bin entanglement [see Fig. \ref{Fig2}]. In Franson interferometry the two time-bin input state passes through an imbalanced interferometer with time delay $\tau$ equal to the time difference between the time bins. This produces 3 different states projections $\{\ket{1},\ket{2},\ket{S}\}$ at the output, where $\ket{S}$ is the superposition state defined as

\begin{equation}
\ket{S}=\frac{1}{\sqrt[]{2}}\left(\ket{1}+e^{i\phi}\ket{2}\right)
\label{Eq3}
\end{equation}

\begin{figure}[t]
\centering\includegraphics[width=\textwidth]{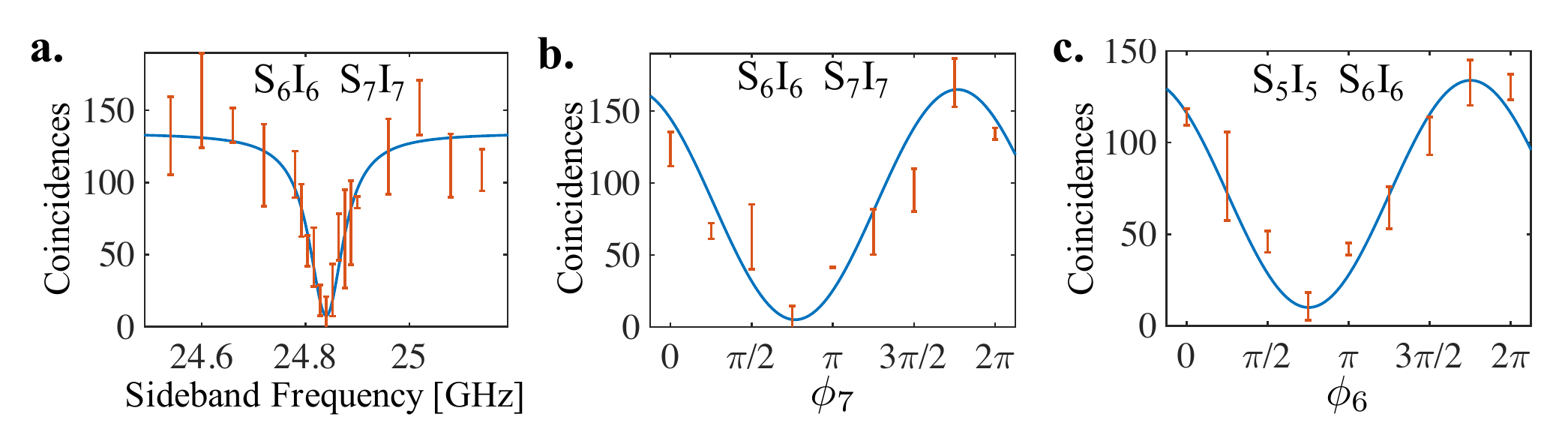}
\caption{(a) Coincidence dip as a function of sideband frequency to maximize the indistinguishability. (b) Coincidences of the S$_6$I$_6$ and S$_7$I$_7$ superposition versus phase applied on S$_7$I$_7$. (c) Coincidences of the S$_5$I$_5$ and S$_6$I$_6$ superposition versus phase applied on S$_6$I$_6$. The coincidences reported are in (a) 20 minutes. and (b), (c) 10 minutes. and after background subtraction. Each data point was measured three times to obtain the standard deviation indicated by the error bars.}
\label{Fig3}
\end{figure}

\begin{figure}[t]
\centering\includegraphics[width=10cm]{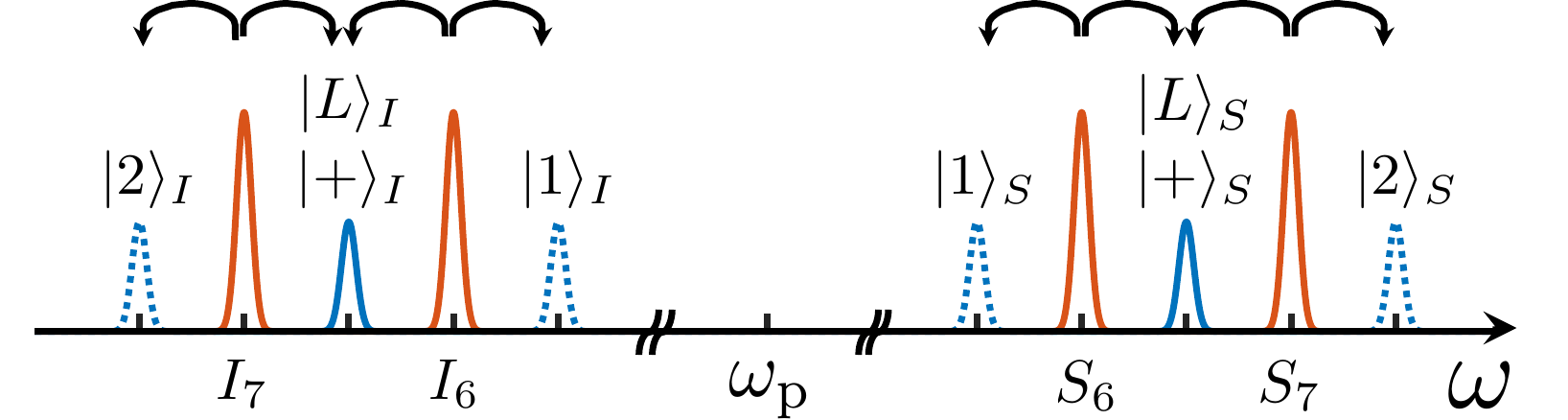}
\caption{Phase modulation scheme for quantum state tomography. Red peaks represent the input signal and idler, each of which is in one of two frequency bins. Blue curves represent projections of signal and idler after the phase modulator (frequency splitter) into three new frequency positions. Solid blue is a projection of the superposition state; dashed blue peaks represent a projection from a single signal or idler frequency bin.}
\label{Fig4}
\end{figure}

\noindent and $\phi$ is a relative phase varied in one of the interferometer arms. In our scheme, we pass a two frequency-bin input state with frequency spacing $\Delta f$ through the phase modulator, which produces upper and lower sidebands at frequency offsets $\pm\Delta f/2$ from each of the parent signals and idlers. In Fig. \ref{Fig2} we label this operation as a ``frequency splitter''. The upper sideband from one parent signal (idler) frequency overlaps with the lower sideband from the other parent signal (idler) frequency. Accordingly, at the output of the frequency splitter, we will have 3 different state projections $\{\ket{1},\ket{2},\ket{S}\}$ where $\ket{S}$ is the superposition state again defined as in Eq. (\ref{Eq3}), but now with $\phi$ corresponding to a phase imposed onto the biphoton by the first pulse shaper prior to the phase modulator. We can apply different relative phases between the parent frequency bins, and therefore the superposition state $\ket{S}$ can have different representations according to Eq. (\ref{Eq3}). We note that unlike Franson interferometry, where phase stabilization is needed, here the phases in our frequency interferometry approach are intrinsically stable.

To be able to measure the optimum frequency overlap and maximize the indistinguishability between different phase modulation sidebands, first we apply a relative phase shift of $\pi$ between S$_6$I$_6$ and S$_7$I$_7$ using pulse shaper 1---inducing a $\pi/2$ phase on both S$_6$ and I$_6$---to create a destructive interference between these two modes. We proceed to measure the coincidences as we sweep the rf frequency to yield a sideband separation from 24.54 to 25.14 GHz. We observe a dip with a maximum visibility of 89\% at 24.84 GHz, as shown in Fig. \ref{Fig3}(a). The full width at half maximum of this dip is measured to be $\sim$100 MHz, similar to the resonance linewidth of the microring. We note that background accidentals were subtracted from the plot in Fig. \ref{Fig3}(a) and subsequent results in the rest of the paper, where the coincidence to accidental ratio was about 2:1. This reduction in coincidence to accidental ratio in the phase measurement experiments compared to the JSI measurement is due to the additional loss that the extra pulse shaper and phase modulator introduce to our biphotons; as a consequence we are forced to use higher pump power and biphoton flux, which reduces the ratio. 

Now that we have superposition of the sidebands, we should be able to observe an interference pattern by changing the relative phases of the comb line pairs. Using the first pulse shaper to vary the phases of S$_7$ and I$_7$ simultaneously, we obtain a sinusoidal interference pattern in the measured coincidences [Fig. \ref{Fig3}(b)]. The resulting visibility of $93\%\pm13\%$ shows strong phase coherence between the comb line pairs S$_6$I$_6$ and S$_7$I$_7$. Following the same procedure but selecting comb line pairs S$_5$I$_5$ and S$_6$I$_6$ and sweeping the phases of S$_6$ and I$_6$ simultaneously, we obtain a visibility of $86\%\pm11\%$ [Fig. \ref{Fig3}(c)]. Our results establish a two-photon interferometry approach for frequency-bin entangled photons that is in close analogy with the Franson (time-imbalanced) interferometer approach widely used for characterization of time-bin entangled photons \cite{franson1989bell}.

\subsection{Quantum state tomography}

We perform quantum state tomography by measuring a complete set of 16 projections of the two-qubit entangled state \cite{james2001measurement,takesue2009implementation} which allows us to estimate the density matrix. We performed coincidence measurements between signal and idler photons in the 16 possible combinations of the states $\{\ket{1},\ket{2},\ket{L},\ket{+}\}$. Here, $\ket{L}$ and  $\ket{+}$ are the superposition states in Eq. (\ref{Eq3}) when $\phi$ is equal to $\pi/2$ and 0, respectively, as shown in Fig. \ref{Fig4}. Because we can make an exact analogy between our approach for projecting frequency-bin qubits and the Franson interferometry approach for projecting time-bin qubits, we can perform quantum state tomography of two-photon frequency-bin qubit states using an exact transcription of the measurement protocol for two-photon time-bin qubit states detailed in \cite{takesue2009implementation}. The estimated real and imaginary parts of the density matrix are shown in Figs. \ref{Fig5}(a) and \ref{Fig5}(b), respectively. (See Appendix B for more details).

\begin{figure}[t]
\centering\includegraphics[width=12cm]{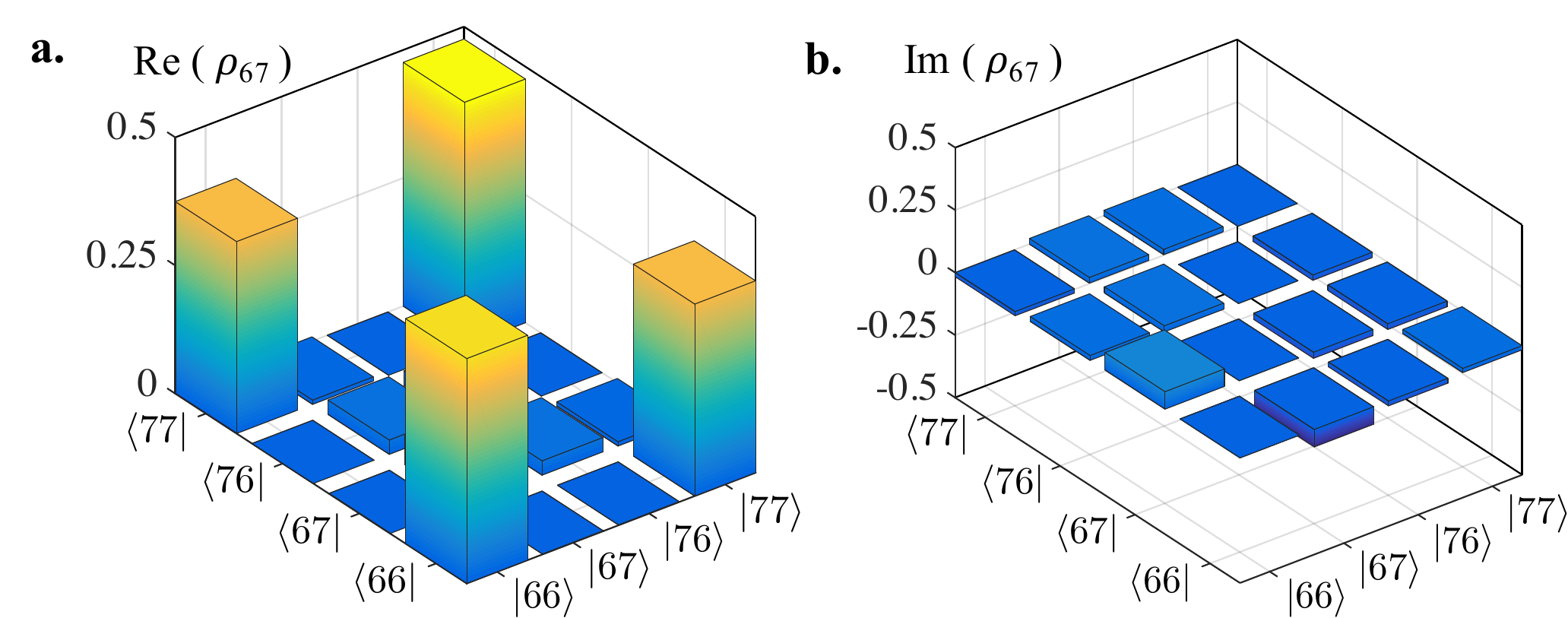}
\caption{(a) Real and (b) imaginary parts of the estimated density matrix for comb line pairs S$_6$I$_6$ and S$_7$I$_7$.}
\label{Fig5}
\end{figure}

To evaluate the amount of entanglement in the measured two-qubit state, we use the Peres-Horodecki criterion \cite{peres1996separability,horodecki1996separability} and calculate an entanglement monotone called negativity. The negativity of a density matrix $\hat{\rho}$ is defined as: $N(\hat{\rho})=\sum_{i=0}^{3}\frac{\lvert\lambda_i\rvert-\lambda_i}{2}$ ,where $\lambda_i$ are the eigenvalues of the partial positive transposed version of $\hat{\rho}$. A two-qubit density matrix is separable iff $N(\hat{\rho})=0$, and $N(\hat{\rho})>0$ signifies entanglement. For a maximally entangled state $N(\hat{\rho})=0.5$, and for the experimentally recovered state given in Appendix B we find $N(\hat{\rho})=0.34$, strongly indicating inseparability.

\subsection{Three-dimensional frequency-bin entanglement}

\begin{figure}[t]
\centering\includegraphics[width=10cm]{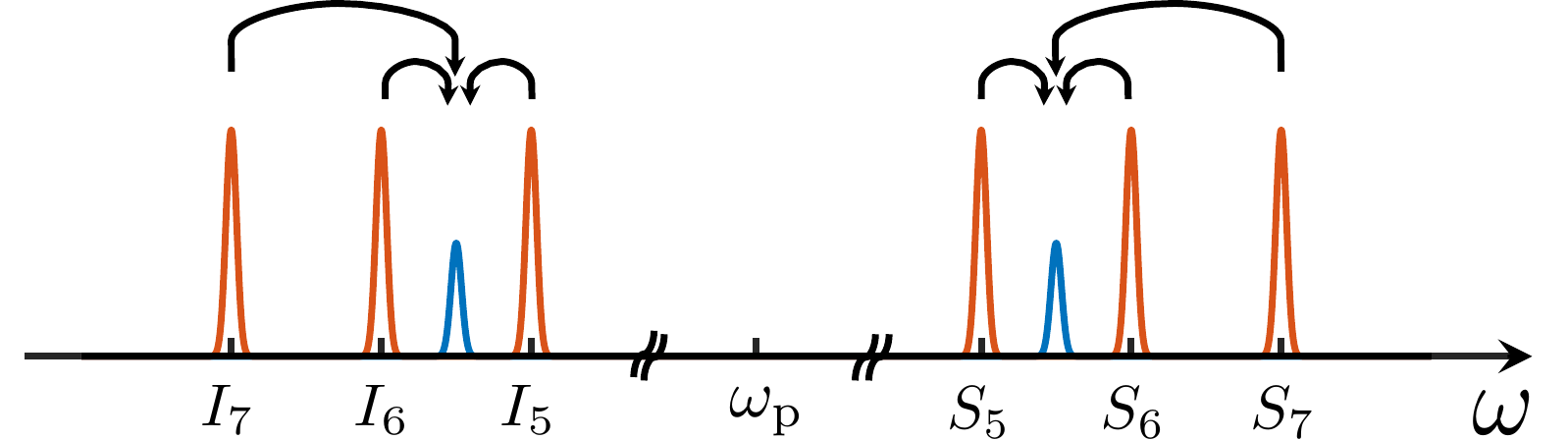}
\caption{Illustration of overlapped phase modulation sidebands for comb line pairs S$_5$I$_5$, S$_6$I$_6$ and S$_7$I$_7$.}
\label{Fig6}
\end{figure}

The results presented so far have been for two-dimensional quantum states. Our observation of strong interference contrast involving comb line pairs S$_5$I$_5$-S$_6$I$_6$ and S$_6$I$_6$-S$_7$I$_7$ individually suggests phase coherence across lines 5, 6 and 7 jointly. For a proof of such high-dimensionality, however, we must examine phase coherence across the selected comb line pairs simultaneously. Here we consider a biphoton state initially made up of three comb line pairs (two entangled qutrits). We use the first pulse shaper to select the comb line pairs S$_5$I$_5$, S$_6$I$_6$ and S$_7$I$_7$; after the phase modulator, we overlap the first sidebands for the 5$^{\textrm{th}}$ and 6$^{\textrm{th}}$ comb line pairs together with the third sideband from the 7$^{\textrm{th}}$ comb line pair. In order to ensure equal mixing weights for all three sidebands, we send a continuous-wave test laser through the modulator and adjust the electrical drive power such that the first and third phase modulation sidebands are equalized, as verified with an optical spectrum analyzer. We also use the first pulse shaper to balance the intensities of the biphoton sideband pairs such that individually they each contribute equal coincidence counts (so the three diagonal terms in the JSI are equal), thereby maximizing the potential Bell inequality violation. Additionally, we compensate for the relative phases on the comb line pairs induced by fiber dispersion. Now we use the second pulse shaper to select the overlapping sidebands from the signal and idler triplets [blue curves in Fig. \ref{Fig6}], which arise from an indistinguishable superposition of contributions from S$_{5}$, S$_{6}$, S$_{7}$ and I$_{5}$, I$_{6}$, I$_{7}$, respectively. Pulse shaper 1 places spectral phases on the signal and idler lines such that the ideal state after the second pulse shaper can be written in the form $\ket{\psi} \propto \ket{5,5}_{SI} + e^{i\left(\phi_S+\phi_I\right)} \ket{6,6}_{SI} + e^{i2\left(\phi_S+\phi_I\right)} \ket{7,7}_{SI}$. Extensive numerical searches \cite{collins2002bell} suggest that the largest violation of the 3-dimensional Bell inequality is realized by measurement bases with the property that the phase applied to the 7$^{\textrm{th}}$ signal and idler should be twice the phase put on the 6$^{\textrm{th}}$ comb line pair \cite{thew2004bell}. Now, by setting the phase parameters $\phi_S$ and $\phi_I$ to appropriate specific values, we construct a three-dimensional CGLMP inequality ($I_3\le2$) adapted from \cite{collins2002bell} and described in detail for time-bin and frequency-bin entangled photons in \cite{thew2004bell,bernhard2014non}, respectively (see Appendix C). We calculate $I_3=2.63\pm0.2$ which surpasses the classical limit of 2 by more than three standard deviations. This shows a phase coherence spanning three comb line modes and validates high dimensional frequency-bin entanglement for our BFC.

\section{Discussion}

While we have demonstrated frequency-bin entanglement for up to 3 dimensions, it is of great interest to extend this scheme to investigate entanglement and reconstruct density matrices for even higher dimensions. Significant improvements to the experimental apparatus that would help in this endeavor can be readily foreseen. The most obvious would be to replace the InGaAs single-photon detectors with superconducting nanowire single-photon detectors (SNSPDs) \cite{kues2017chip}. Such detectors can provide quantum efficiencies >80\% and dead times on the order of 100 ns. Therefore, with an upgrade to SNSPD detectors, the count rate in our experiments would be increased by an impressive factor of $\sim$100 (a factor of 10 from the improved efficiency of two detectors and another factor of 10 from reduced dead time). Furthermore, since the SNSPDs have only <100 dark counts/sec, three orders of magnitude better than our current detectors, the background counts should be strongly reduced. These factors would allow us to reach significantly higher visibilities even without background subtraction. Additional enhancement is possible using parallel detection. Commercial pulse shaper technology supports programmable routing of different frequency channels to more than one dozen different output fibers. As an example, in our two qubit frequency-bin quantum state tomography scheme, if all six output frequency channels (three for signal, three for idler) were routed to six different output fibers connected to parallel SNSPDs and multi-channel timing electronics, it should be possible to reduce the number of measurement cycles from 36 to 4, providing a further factor of 9 improvement. This is similar to the measurement speed-up reported in two qubit time-bin quantum state tomography \cite{takesue2009implementation} where $\ket{1}$, $\ket{2}$ and $\ket{S}$ photons are time resolved and recorded in the same measurement cycle. Parallel detection schemes would also be beneficial for higher dimensional experiments. Although the phase modulator spreads energy from individual signal or idlers into multiple sidebands, only one sideband per signal or idler is used in the current experiments; energy spread to the unused sidebands is lost. For example, our frequency conversion efficiency using the PM is currently about 30\% when we optimize photon transfer to the first sideband and about 15\% when we transfer to both first and third sidebands. With parallel detection with a sufficient number of detectors, multiple sidebands lying between original biphoton comb lines could be used, substantially mitigating unnecessary loss and opening the door to stronger phase modulation to construct superpositions of a larger number of frequency bins. Tailoring the rf waveform driving the phase modulator could also contribute to improving efficiency \cite{lukens2017frequency}.

Algorithmic innovations may also aid in quantifying frequency-bin entanglements over larger subspaces. The number of measurements required to fully reconstruct the density matrix through quantum state tomography grows rapidly with increased dimensionality.  New methods which provide bounds on high dimensional entanglement based on measurements that only partially sample the density matrix \cite{martin2017quantifying} should provide a more favorable scaling.

Finally, we note that while the on-chip biphoton source is fairly efficient, we incur a large loss of about 15 dB simply due to the insertion loss of the discrete off-chip components (phase modulator and two pulse shapers). It will be interesting to investigate the potential for reducing this loss through photonic integration.  Quantum photonic chips based on arrays of interferometers are now an active area of research \cite{politi2008silica}. For studies of frequency bin entanglement, a more appropriate architecture could include the microresonator biphoton source and on-chip phase modulators and pulse shapers. The pulse shapers, for example, could be constructed from thermally tunable arrays of microring resonators, which have been demonstrated with both spectral amplitude \cite{khan2010ultrabroad} and spectral phase shaping functionalities \cite{agarwal2006fully} for applications in rf photonic and optical signal processing.

\section{Conclusion}

In summary, our research offers a scalable integrated platform to generate high dimensional photonic states in a superposition of different frequency bins. Due to its robustness and weak interaction with the environment, the frequency degree of freedom in photonic states is a potential candidate to move this research towards experimental realization of high dimensional quantum computing protocols. The use of these high dimensional entangled states offers a clear path to having more complex quantum circuits within reach, as well as denser information encoding \cite{barreiro2008beating,lanyon2009simplifying}.

\section*{Appendices:}

\section*{\textit{A. Experimental details}}

Our scheme for characterizing the frequency bin entanglement is based on commercial instrumentation such as pulse shapers, phase modulators, and single photon detectors, all of which are fiber pigtailed and compatible with operation in the lightwave C band. The microring resonator is formed from waveguides with dimensions 1.6 $\mu$m wide by 870 nm thick fabricated in a SiN film. In- and out-coupling to the SiN chips are performed with lensed fibers. U-grooves etched into the chip [see Fig. \ref{Fig1}(a)] provide support points that enhance the stability of the coupling \cite{xuan2016high}. Interference filters [not depicted in Fig. \ref{Fig1}(d)] follow the microring and strongly attenuate the pump line; sideband pairs S$_1$I$_1$ and S$_2$I$_2$ are also attenuated in the process. Pulse shapers 1 and 2 (Finisar WaveShaper models 1000S and 4000S, respectively) allow us to perform programmable filtering with 10 GHz resolution and 1 GHz addressability over the wavelength ranges 1527.4--1567.5 nm and 1527.4--1600.8 nm, respectively. Pulse shaper 2 also supports programmable frequency selective routing to four fiber output ports (only two are used in the current experiments). Based on the availability of phase modulators (lithium niobate integrated optic modulators from EOSpace), we used a 20-GHz bandwidth modulator for the frequency qubit measurements of Fig. \ref{Fig3}. We modulated with an rf frequency of 12.4 GHz and used the $\pm$2 sidebands corresponding to $\pm$24.8 GHz frequency offset to get the frequency overlapped superposition. For the frequency qutrit measurements of Fig. \ref{Fig6}, a higher (40 GHz) bandwidth modulator was available, allowing us to modulate directly at 24.8 GHz. An advantage of using a relatively large microresonator with correspondingly small (49.6 GHz) free spectral range is its relatively good match with practical rf modulation frequencies; by working with low-order modulation sidebands, we are able to shift a relatively large fraction of the signal and idler power into the sidebands used for superposition. Coincidences are measured using a pair of InGaAs single-photon avalanche diodes (Aurea Technology) connected to a two-channel time-to-digital converter module (PicoQuant HydraHarp). The detectors have specified 25\% quantum efficiency, 1000 ns dead time, and 10$^5$ dark counts/sec with a gate frequency of 1.25 MHz.

Using this experimental setup, we first find the best rf drive frequency for maximum indistinguishability between the frequency bins S$_6$I$_6$ and S$_7$I$_7$. In this process, we program pulse shaper 1 (taking into account the estimated dispersion of fiber leads) for a phase difference of $\pi$ between S$_6$I$_6$ and S$_7$I$_7$; this condition is expected to yield destructive interference and the minimum number of coincidences after the rf frequency is optimized [see Fig. \ref{Fig3}(a)]. To obtain a complete interference pattern, we sweep the phase of S$_7$I$_7$ over a range of 2$\pi$ [Fig. \ref{Fig3}(b)], recording coincidences for ten minutes at each phase point. To extract the visibilities, we use the expression $V=\left(C_{\textrm{max}}-C_{\textrm{min}}\right)/\left(C_{\textrm{max}}+C_{\textrm{min}}\right)$, where $C_{\textrm{max}}$ and $C_{\textrm{min}}$ correspond to the phase settings where the maximum and minimum coincidences are expected. This procedure for estimating the visibility is repeated three times to yield an average and standard deviation for the visibility estimate.

The effect of dispersion due to fiber leads can be seen in Figs. \ref{Fig3}(b) and \ref{Fig3}(c) as a shift in the sinusoidal interference patterns. Without dispersion, the maxima of the interference patterns should be at zero phase, but we can see a shift of $\sim\pi/4$ in the measured interference patterns. From this shift, the amount of standard single mode fiber in our setup can be estimated ($\sim$35 m). We use this calculated fiber length to compensate for dispersion (by programming the pulse shaper for additional phase to offset the frequency dependent phase from the dispersion) in our measurement of the three-dimensional CGLMP inequality described in section 2.4.

\section*{\textit{B. Density matrix reconstruction}}

The measurement protocol and coincidence count data for the quantum state tomography (Section 2.3) are given in Table\ref{tab:1}. Table \ref{tab:1} may be understood as follows. Since the two-qubit density matrix is $4\times4$, we require a complete set of 16 projections $\ket{\Psi_{\nu}} (\nu=1:16)$ , written in terms of its basis coefficients $(\bra{11}\Psi_{\nu}\rangle, \bra{12}\Psi_{\nu}\rangle, \bra{21}\Psi_{\nu}\rangle, \bra{22}\Psi_{\nu}\rangle)$ . We perform these projections by acquiring data in four different phase configurations $(\phi_S,\phi_I)=\{(0,0),(0,\pi/2),(\pi/2,0),(\pi/2,\pi/2)\}$ , columns 5-8. Here, in performing tomography on the S$_6$I$_6$-S$_7$I$_7$ qubit pair, $\phi_S$ and $\phi_I$ are the signal and idler phases applied to the 7$^{\textrm{th}}$ comb line pair via pulse shaper 1 in the experimental setup shown in Fig. \ref{Fig1}(d). For each projection columns 2 and 3 specify which signal and idler frequency channel are routed to the respective single photon detectors. Referring to Fig. \ref{Fig4}, $\ket{1}$ and $\ket{2}$ in columns 2 and 3 correspond to unique frequency channels, whereas $\ket{+}$ and $\ket{L}$ are both sideband superpositions measured when the same physical frequency channel is routed for detection. Therefore, an entry in column 2 of $\ket{+}$ or $\ket{L}$ signifies both routing of the signal superposition frequency channel for detection and application of the appropriate phase to the 7$^{\textrm{th}}$ signal line (0 phase for $\ket{+}$ , data reported in column 5 or 6; $\pi/2$ phase for $\ket{L}$, data reported in column 7 or 8). An entry in column 3 of $\ket{+}$ or $\ket{L}$ has similar meaning, but refers to the idler superposition frequency channel (data in column 5 or 7 for $\ket{+}$, column 6 or 8 for $\ket{L}$). As an example, for $\ket{\Psi_8}$ we have $(\phi_S,\phi_I)=(\pi/2,0)$, column 7, and we obtain:

\begin{equation}
\begin{gathered}
\ket{\Psi_8} =  \frac{1}{\sqrt[]{2}}\left(\ket{1}_S+e^{i\phi_S}\ket{2}_S\right).\frac{1}{\sqrt[]{2}}\left(\ket{1}_I+e^{i\phi_I}\ket{2}_I\right)\\
= \frac{1}{2}\ket{1,1}_{SI}+\frac{1}{2}\ket{1,2}_{SI}+\frac{i}{2}\ket{2,1}_{SI}+\frac{i}{2}\ket{2,2}_{SI}\\
= \left(\frac{1}{2},\frac{1}{2},\frac{i}{2},\frac{i}{2}\right)
\end{gathered}
\label{eq:4}
\end{equation}
In this notation $\ket{x,y}_{SI}=\ket{x}_{S}\ket{y}_{I}$, in which signal and idler photons are in frequency bins $x$ and $y$, respectively.

Also, as explained in \cite{takesue2009implementation}, for each of the signal and idler photons, measurement in a nonsuperposition basis ($\ket{1}$ or $\ket{2}$) involves a factor of two loss relative to measurement in the superposition channel. This is understood in the time-bin case as the loss incurred at the output beam splitter of the interferometer, since for nonsuperposition bases, half of the photons go to the unused output port. For the superposition cases, with constructive interference such loss is avoided.  The same argument holds in our frequency-bin approach. These factors of two that arise for each of signal and idler are accounted for by noting for projections such as $\ket{\Psi_1}=\ket{11}$, which incur a factor of four loss, coincidences may be measured for each of the four phase configurations. The corresponding coincidence counts are listed in columns 5 to 8 and are added to give a total coincidence count (column 9). Likewise, projections such as $\ket{\Psi_6}=\ket{1+}$ incur a factor of two loss but may be measured in two phase configurations, and projections such as $\ket{\Psi_7}=\ket{++}$ incur no extra loss but are measured in only a single-phase configuration. Overall, 36 independent measurements are performed, and the total number of coincidence counts obtained by adding the entries in columns 5-8 (column 9, $n_{\nu}$) provides the correct normalization across the different projections. 

As in \cite{james2001measurement,takesue2009implementation}, we perform a maximum likelihood estimate to obtain the density matrix that best fits our projection measurement data (the $n_{\nu}$) while satisfying the requirement for a physical density matrix that the eigenvalues lie in the interval [0,1]. This estimation is calculated using the minimization of the following likelihood function:

\begin{equation}
\mathcal{L}= \sum\limits_{\nu=1}^{16}\frac{\big(C\bra{\Psi_\nu}\hat{\rho}\ket{\Psi_\nu}-n_\nu\big)^2}{2C\bra{\Psi_\nu}\hat{\rho}\ket{\Psi_\nu}}
\label{eq:5}
\end{equation}

\begin{table}[H]
\caption [Projection measurements for frequency-bin density matrix estimation.]{Projection measurements for frequency-bin density matrix estimation.  For each measurement coincidences were acquired over a 10-minute period.  A dash (-) indicates that the phase setting indicated by the respective column is not involved in the projection measurement indicated by the respective row; hence coincidence counts were not obtained.} 
\label{tab:1}
\def\arraystretch{2}
\begin{center}
  \begin{tabular}{|c|c|c|c|c|c|c|c|c|}
    \hline
    & Signal & Idler & &\multicolumn{4}{c|}{$\left(\phi_S,\phi_I\right)$} &   \\ \cline{5-8}
    $\nu$ & Photon & Photon & $\ket{\Psi_\nu}$ & $\left(0,0\right)$ & $\left(0,\frac{\pi}{2}\right)$ & $\left(\frac{\pi}{2},0\right)$ & $\left(\frac{\pi}{2},\frac{\pi}{2}\right)$ & $n_\nu$ \\ \hline
    1 & $\ket{1}$ & $\ket{1}$ & $(1,0,0,0)$ & 36 & 40 & 36 & 41 & 153\\ \hline
    2 & $\ket{1}$ & $\ket{2}$ & $(0,1,0,0)$ & 9 & 8 & 0 & 0 & 17\\ \hline
    3 & $\ket{2}$ & $\ket{1}$ & $(0,0,1,0)$ & 0 & 0 & 0 & 7 & 7\\ \hline
    4 & $\ket{2}$ & $\ket{2}$ & $(0,0,0,1)$ & 47 & 29 & 44 & 31 & 151\\ \hline
    5 & $\ket{2}$ & $\ket{+}$ & $\left(0,0,\frac{1}{\sqrt[]{2}},\frac{1}{\sqrt[]{2}}\right)$ & 26 & - & 40 & - & 66\\ \hline   
    6 & $\ket{1}$ & $\ket{+}$ & $\left(\frac{1}{\sqrt[]{2}},\frac{1}{\sqrt[]{2}},0,0\right)$ & 47 & - & 22 & - & 69\\ \hline 
    7 & $\ket{+}$ & $\ket{+}$ & $\left(\frac{1}{2},\frac{1}{2},\frac{1}{2},\frac{1}{2}\right)$ & 146 & - & - & - & 146\\ \hline 
	8 & $\ket{L}$ & $\ket{+}$ & $\left(\frac{1}{2},\frac{1}{2},\frac{i}{2},\frac{i}{2}\right)$ & - & - & 71 & - & 71 \\ \hline
    9 & $\ket{L}$ & $\ket{1}$ & $\left(\frac{1}{\sqrt[]{2}},0,\frac{i}{\sqrt[]{2}},0\right)$ & - & - & 14 & 57 & 71\\ \hline
    10 & $\ket{L}$ & $\ket{2}$ & $\left(0,\frac{1}{\sqrt[]{2}},0,\frac{i}{\sqrt[]{2}}\right)$ & - & - & 26 & 44 & 70\\ \hline
    11 & $\ket{L}$ & $\ket{L}$ & $\left(\frac{1}{2},\frac{i}{2},\frac{i}{2},\frac{-1}{2}\right)$ & - & - & - & 4 & 4\\ \hline
    12 & $\ket{1}$ & $\ket{L}$ & $\left(\frac{1}{\sqrt[]{2}},\frac{i}{\sqrt[]{2}},0,0\right)$ & - & 21 & - & 29 & 50\\ \hline
    13 & $\ket{2}$ & $\ket{L}$ & $\left(0,0,\frac{1}{\sqrt[]{2}},\frac{i}{\sqrt[]{2}}\right)$ & - & 44 & - & 31 & 75\\ \hline
    14 & $\ket{+}$ & $\ket{L}$ & $\left(\frac{1}{2},\frac{i}{2},\frac{1}{2},\frac{i}{2}\right)$ & - & 62 & - & - & 62\\ \hline
    15 & $\ket{+}$ & $\ket{1}$ & $\left(\frac{1}{\sqrt[]{2}},0,\frac{1}{\sqrt[]{2}},0\right)$ & 16 & 29 & - & - & 45\\ \hline 
    16 & $\ket{+}$ & $\ket{2}$ & $\left(0,\frac{1}{\sqrt[]{2}},0,\frac{1}{\sqrt[]{2}}\right)$ & 49 & 32 & - & - & 81\\ \hline     
  \end{tabular}
\end{center}
\end{table}

\noindent where $C$ is the normalization constant defined by:

\begin{equation}
C=\sum\limits_{\nu=1}^{4} n_\nu
\label{eq:6}
\end{equation}

As a result of this optimization, we found the following physical density matrix:

\begin{equation}
\hat{\rho}=
\begin{bmatrix}
0.4388+0.0000i & -0.0115-0.0699i & -0.0721-0.0193i & 0.3745+0.0166i \\
-0.0115+0.0699i & 0.0574+0.0000i & 0.0279-0.0244i & 0.0084-0.0227i \\
-0.0721+0.0193i & 0.0279+0.0244i & 0.0281+0.0000i & -0.0280-0.0211i \\
0.3745-0.0166i & 0.0084+0.0227i & -0.0280+0.0211i & 0.4757+0.0000i
\end{bmatrix}
\label{eq:7}
\end{equation}
\section*{\textit{C. CGLMP inequality for two qutrits}}

In this section, following \cite{thew2004bell,bernhard2014non}, we describe how we evaluate the CGLMP inequality for two entangled frequency-bin qutrits. As described in the main text, we measure coincidences between signal and idler frequency channels selected to represent superpositions from three parent signal and idler frequencies, respectively. Reference \cite{thew2004bell} evaluated the three-dimensional Bell's inequality for time-bin entangled qutrit states using three-arm interferometers coupled to three different output ports via a $3\times3$ splitter. They constructed a CGLMP inequality expressed in a form equivalent to the following:

\begin{equation}
 \begin{split}
  I_3= 3\left[P^{11}(0,0)+P^{21}(0,1)+P^{22}(0,0)+P^{12}(0,0)\right]\\
  -3\left[P^{11}(0,1)+P^{21}(0,0)+P^{22}(0,1)+P^{12}(1,0)\right]\leq 2
 \end{split}
\label{eq:8}
\end{equation}
where $P^{xy}(a,b)$ is the probability of getting a coincidence count between detector $a$ on the signal and detector $b$ on the idler side, using the measurement basis $A_x$ and $B_y$ for signal and idler photons, respectively. We note that the original form of the 3-dimensional Bell inequality consists of 24 total measurement probabilities \cite{collins2002bell}; the reduction to 8 terms [Eq. (\ref{eq:8})], however, is valid under the assumption of an unbiased $3\times3$ splitter and an input quantum state containing sufficient symmetries. In particular, as we show below, the above reduction holds for a density matrix $\hat{\rho}$ taken to be the incoherent mixture of a maximally entangled state and white noise \cite{thew2004bell}. Such an assumption is physically reasonable and common in visibility-based Bell-violation tests. For the time-bin case, the measurement bases correspond to the sets of phases applied to short, medium, and long interferometer arms. The particular sets of phases used are [$A_1=(0,0,0), A_2=(0,\pi/3,2\pi/3)$] for the signal and [$B_1=(0,\pi/6,\pi/3), B_2=(0,-\pi/6,-\pi/3)$]  for the idler. These choices of phases have been shown to give the largest violation of the CGLMP inequality for a maximally entangled state \cite{thew2004bell}. In the classical picture, if the signal and idler are two independent systems, meaning a measurement on the signal does not affect the idler, and vice versa, then $I_3 \leq 2$. However, for an entangled state this classical limit may be violated, and with the set of phases specified, a maximum violation $I_{3\hspace{1pt}\textrm{max}}=2.872$ is predicted for a maximally entangled state.

In our frequency bin case, the different measurement bases are constructed by putting different sets of phases on different comb line triplets. For example, for a triplet consisting of comb lines 5-7 as in our experiment, for signal measurement basis $A_2$ we would place phases $(0,\pi/3,2\pi/3)$ on signal lines 5, 6, and 7, respectively. However, unlike the  $3\times3$ beam splitter case, we have only a single detector each for signal and idler. This can be accounted for by imposing additional phases on the comb lines according to the equivalent transfer function of the $3\times3$ beam splitter \cite{zukowski1997realizable,kaszlikowski2002clauser}. Equalizing the power in the $\pm1$ and $\pm3$ phase modulation sidebands gives us the ability to perform an unbiased  beam splitter in frequency, thereby satisfying one of the key assumptions behind the reduced form [Eq. (\ref{eq:8})]. The phases are chosen from $\{0,-2\pi/3,2\pi/3\}$ according to which ``beam splitter output'' is involved in the projection that we are mapping from the three-output time-bin case to our one-output frequency-bin case \cite{thew2004bell,bernhard2014non}. In this way, we adapt the CGLMP inequality for time-bin entangled photons to our frequency bins by applying different sets of phases to our comb lines \cite{bernhard2014non}. For our experiment involving comb lines 5, 6, and 7, the phases applied to signal and idler lines $k$ are given by:

\begin{equation}
\Phi_{S_k}^{x}(a) =(k-5)\phi_{S}^{x}(a)= \frac{2\pi}{3}(k-5)\left(a+\alpha_x\right)
\label{eq:9}
\end{equation}

\begin{equation}
\Phi_{I_k}^{y}(b) = (k-5)\phi_{I}^{y}(b)=\frac{2\pi}{3}(k-5)\left(-b+\beta_y\right)
\label{eq:10}
\end{equation}

\noindent Here, $\Phi_{S_k}^{x}(a)$ and $\Phi_{I_k}^{y}(b)$ are the phases applied to the $k^{\textrm{th}}$ signal and idler frequency bin, respectively, expressed in terms of fundamental phases $\phi_{S}^{x}(a)$ and $\phi_{I}^{y}(b)$ for each basis choice $x$ for signal and $y$ for idler; the $a,b = \{0,1,2\}$ correspond to the output channel used in the $3\times3$ splitter version of the projection. The $\alpha_x$ and $\beta_y$ parameters relate to the measurement bases and are chosen as $\alpha_1=0$, $\alpha_2=1/2$, $\beta_1=1/4$ and $\beta_2=-1/4$ . These correspond to the measurement bases $A_x$ and $B_y$ discussed above and yield phase triplets $A_x=(0,(2\pi/3)\hspace{2pt}\alpha_x,(4\pi/3)\hspace{2pt}\alpha_x)$ and $B_y=(0,(2\pi/3)\hspace{2pt}\beta_y,(4\pi/3)\hspace{2pt}\beta_y)$. These are modified by the addition of phase triplets $(0,(2\pi/3)\hspace{2pt}a,(4\pi/3)\hspace{2pt}a)$ and $(0,(-2\pi/3)\hspace{2pt}b,(-4\pi/3)\hspace{2pt}b)$  to signal and idler, respectively, in accord with the $a$ and $b$ parameters. 

As our quantum state, we assume a density matrix of the form

\begin{equation}
\hat{\rho}=\lambda \ket{\psi}\bra{\psi}+(1-\lambda)\hat{\rho}_N
\label{eq:11}
\end{equation}

\noindent with $0 \leq \lambda \leq 1$ , where $\ket{\psi}$ is the maximally entangled state represented as:

\begin{equation}
\ket{\psi} =\frac{1}{\sqrt{3}}\big[ \ket{5,5}_{SI} + \ket{6,6}_{SI}+\ket{7,7}_{SI}\big]
\label{eq:Sup12}
\end{equation}

\noindent and $\hat{\rho}_N$ is our particular noise model, taken to be symmetric, or white:

\begin{equation}
\begin{gathered}
\hat{\rho}_N=\frac{1}{9}\big[ \ket{5,5}\bra{5,5}_{SI}+\ket{5,6}\bra{5,6}_{SI}+\ket{5,7}\bra{5,7}_{SI}+\ket{6,5}\bra{6,5}_{SI}+\ket{6,6}\bra{6,6}_{SI} \\
+\ket{6,7}\bra{6,7}_{SI}+\ket{7,5}\bra{7,5}_{SI}+\ket{7,6}\bra{7,6}_{SI}+\ket{7,7}\bra{7,7}_{SI}\big]
\end{gathered}
\label{eq:13}
\end{equation}

\noindent Following the discussion surrounding Eqs. (\ref{eq:9}) and (\ref{eq:10}), the projective measurements done on each photon are:

\begin{equation}
\hat{\Pi}_S^x(a)=\frac{1}{3}\left[ \ket{5}_{S} + e^{i\phi_S^x(a)}\ket{6}_{S}+e^{i2\phi_S^x(a)}\ket{7}_{S}\right]\left[ \bra{5}_{S} + e^{-i\phi_S^x(a)}\bra{6}_{S}+e^{-i2\phi_S^x(a)}\bra{7}_{S}\right]
\label{eq:14}
\end{equation}

\begin{equation}
\hat{\Pi}_I^y(b)=\frac{1}{3}\left[ \ket{5}_{I} + e^{i\phi_I^y(b)}\ket{6}_{I}+e^{i2\phi_I^y(b)}\ket{7}_{I}\right]\left[ \bra{5}_{I} + e^{-i\phi_I^y(b)}\bra{6}_{I}+e^{-i2\phi_I^y(b)}\bra{7}_{I}\right]
\label{eq:15}
\end{equation}

\noindent Therefore, the probabilities measured are given by:

\begin{equation}
\begin{gathered}
P^{xy}(a,b)=\textrm{Tr}\left\{\hat{\rho}\hspace{2pt}\hat{\Pi}_S^x(a)\otimes\hat{\Pi}_I^y(b)\right\}\\
= \lambda \big\langle \psi \big| \hat{\Pi}_S^x(a)\otimes\hat{\Pi}_I^y(b) \big| \psi \big\rangle + \frac{1-\lambda}{9}\sum_{m=5}^7 \sum_{n=5}^7 \big\langle mn \big| \hat{\Pi}_S^x(a)\otimes\hat{\Pi}_I^y(b)\big| mn \big\rangle_{SI}
\end{gathered}
\label{eq:16}
\end{equation}

\noindent The noise matrix elements all evaluate to

\begin{equation}
\big\langle mn \big| \hat{\Pi}_S^x(a)\otimes\hat{\Pi}_I^y(b) \big| mn \big\rangle_{SI}=\frac{1}{9}
\label{eq:17}
\end{equation}

\noindent  and the first term in Eq. (\ref{eq:16}) reduces to

\begin{equation}
\big\langle \psi \big| \hat{\Pi}_S^x(a)\otimes\hat{\Pi}_I^y(b) \big| \psi \big\rangle= \frac{1}{27}\left|1+e^{i\left[\phi_S^x(a)+\phi_I^y(b)\right]}+e^{i2\left[\phi_S^x(a)+\phi_I^y(b)\right]}\right|^2
\label{eq:18}
\end{equation}

\noindent Combined, Eqs. (\ref{eq:17}) and (\ref{eq:18}) justify the simplification from a full 24-term Bell parameter to the 8-term $I_3$ in Eq. (\ref{eq:8}), which is based on symmetries in the combinations of outcomes $a$ and $b$. The noise terms show no dependence on $a$ and $b$ [Eq. (\ref{eq:17})], while the pure state contribution [Eq. (\ref{eq:18})] varies only via the difference $a-b$, modulo 3. Thus, under our particular noise model, we only need to obtain 8 probability estimates. This model is consistent with the measured JSI, which shows a roughly constant background on the off-diagonal terms within the two-qutrit subspace. (We note that a Bell test with no such symmetry assumptions would be possible by testing all 24 projections separately.)

\begin{table}[H]
\caption{Parameters for evaluations of the CGLMP inequality. The coincidences were measured in 10-minute spans; measurements were done three times to obtain standard deviations. To achieve the maximum and minimum number of coincidences, the phases of $\phi_S^x(a)=\phi_I^y(b)=0$  and $\phi_S^x(a)=\phi_I^y(b)=\pi/3$ were put on the biphotons, respectively. To calculate each of the probabilities that appear in Eq. (\ref{eq:8}), the corresponding coincidence counts have to be divided by the maximum number of coincidences $P_{\textrm{max}}(0,0)$.} \label{tab:2}
\begin{center}
\def\arraystretch{1.2}
  \begin{tabular}{|c|c|c|c|c|c|c|c|}
    \hline
   Term  & $x$ & $y$ & $a$ & $b$ & $\phi_S^x(a)$ & $\phi_I^y(b)$ & Coincidences\\ \hline
    $P^{11}(0,0)$ & 1 & 1 & 0 & 0 & 0 & $\pi/6$ & 150$\pm$10\\ \hline
     $P^{21}(0,1)$ & 2 & 1 & 0 & 1 & $\pi/3$ & $-\pi/2$ & 141$\pm$23\\ \hline
    $P^{22}(0,0)$ & 2 & 2 & 0 & 0 & $\pi/3$ & $-\pi/6$ & 152$\pm$21\\ \hline
     $P^{12}(0,0)$ & 1 & 2 & 0 & 0 & 0 & $-\pi/6$ & 146$\pm$16\\ \hline
    $P^{11}(0,1)$ & 1 & 1 & 0 & 1 & 0 & $-\pi/2$ & 54$\pm$4\\ \hline
    $P^{21}(0,0)$ & 2 & 1 & 0 & 0 & $\pi/3$ & $\pi/6$ & 33$\pm$6\\ \hline
     $P^{22}(0,1)$ & 2 & 2 & 0 & 1 & $\pi/3$ & $-5\pi/6$ & 49$\pm$12\\ \hline
     $P^{12}(1,0)$ & 1 & 2 & 1 & 0 & $2\pi/3$ & $-\pi/6$ & 32$\pm$10\\ \hline
     $P_{\textrm{max}}(0,0)$ & - & - & 0 & 0 & 0 & 0 & 160$\pm$18\\ \hline
     $P_{\textrm{min}}(0,0)$ & - & - & 0 & 0 & $\pi/3$ & $\pi/3$ & 15$\pm$13\\ \hline
  \end{tabular}
\end{center}
\end{table}

\noindent In the Table 2, the first column corresponds to the individual terms in Eq. (\ref{eq:8}). Columns 6 and 7 evaluate Eqs. (\ref{eq:9}) and (\ref{eq:10}) to obtain the signal and idler phase parameters $\phi_S^x(a)$ and $\phi_I^y(b)$. Our coincidence data are given in column 8. We calculate $I_3=2.63 \pm 0.2$ which is more than three standard deviations away from the classical limit and indicates three-dimensional frequency-bin entanglement.

\section*{Funding}
National Science Foundation (NSF) (ECCS-1407620), DARPA PULSE Program (W31P40-13-1-0018), Oak Ridge National Laboratory (ORNL).

\section*{Acknowledgment}
We acknowledge Allison L. Rice for designing the graphics of the experimental setup in Fig. 1. J.A.J. acknowledges support by Colciencias and Fulbright Colombia. A portion of this work was performed at Oak Ridge National Laboratory, operated by UT-Battelle for the U.S. Department of Energy under contract no. DE-AC05-00OR22725.

\end{document}